\documentclass[fleqn,twoside]{article}
\usepackage{espcrc2}

\usepackage{graphicx}
\usepackage[figuresright]{rotating}
\usepackage{amsmath}
\usepackage{amssymb}

\newcommand{\eVdist}{\kern-0.06667em}
\newcommand{\gev}{{\,\text{Ge}\eVdist\text{V\/}}}
\newcommand{\pb}{\,\text{pb}}
\newcommand{\nb}{\,\text{nb}}
\newcommand{\NC}{{\rm NC}}
\newcommand{\CC}{{\rm CC}}
\newcommand{\syst}{{\rm sys}}
\newcommand{\PDF}{\text{PDF}}
\newcommand{\stat}{\text{stat}}

\title{Structure Function Results from ZEUS}

\author{A. Kappes\address[MCSD]{Physikalisches Institut der Universit\"at Bonn,
        Nu\ss allee 12, D-53115 Bonn}
      for the ZEUS Collaboration}

\begin{document}

\mathindent=0pt
\setlength{\intextsep}{15pt}
\setlength{\textfloatsep}{10pt}
\setlength{\floatsep}{10pt}
\renewcommand{\topfraction}{1.}
\setcounter{totalnumber}{4}
\setcounter{topnumber}{2}

\begin{abstract}
  This contribution presents recent ZEUS results on proton structure
  functions at HERA. The inclusive $\phi(1020)$-meson cross section
  was measured, and it was used to determine the $s$-quark content of
  the proton. The structure function $F_2$ was extracted using
  initial-state radiative events.  Neutral and charged current cross
  sections were used to extract the structure function $xF_3$ and
  measure the mass of the $W$ boson, respectively. A NLO QCD fit to
  ZEUS data and fixed target cross sections was employed to determine
  the parton density functions of the quarks and of the gluon inside
  the proton. \vspace{1pc}
\end{abstract}

\maketitle

\section{\boldmath Deep inelastic $ep$ scattering (DIS)}
The Hadron-Elektron-Ringanlage HERA at DESY in Hamburg accelerates
electrons/positrons and protons to energies of $E_e = 27.5 \gev$ and
$E_p = 920 \gev$ ($900 \gev$ before 1998), yielding a center-of-mass
energy, $\sqrt{s}$, of $318 \gev$ ($300 \gev$ before 1998). In DIS
events the interaction between the two particles can either occur via
a neutral (NC, $\gamma$ or $Z$) or a charged current (CC, $W$)
process. The reaction can be described by two of the three variables
$Q^2$, $x$-Bjorken and $y$. Here, $Q^2$ is the negative of the squared
4-momentum transfer between lepton and proton and $y$ is the
inelasticity.

The NC cross section can be written as 
\begin{eqnarray*}
  \frac{d^2 \sigma({e^\pm p})} {dx\,dQ^2} 
  = \frac{2 \pi \alpha^2}{x\, Q^4} \left[ Y_+
    {F_2}^\NC \mp Y_- {x F_3}^\NC - y^2 F_L^\NC \right] \\[2mm]
  (F_2^\NC,xF_3^\NC) = x \sum_{q=u \dots b} (A_f,B_f) \left[ (q +
    \bar{q},q - \bar{q})  \right] , \mbox{\hspace{5mm}} 
\end{eqnarray*}
where $q$ and $\bar{q}$ are the parton density functions (PDFs) of
quarks and antiquarks and $Y_\pm = 1 \pm (1-y)^2$. $A_f$ and $B_f$
depend on $Q^2$ only and contain the vector and axial-vector couplings
as well as the propagator terms. The reduced cross section
is defined as $\tilde{\sigma}^\NC = \frac{d^2 \sigma^\NC} {dx\,dQ^2}
\times \frac{x Q^4}{2 \pi \alpha^2} \frac{1}{Y_+}$.

The CC cross section is similar to that of NC
\begin{eqnarray*}
\frac{d^2 \sigma (e^- p)} {dx\,dQ^2}
  = \frac{G_F^2}{4 \pi x} \frac{M_W^4}{(Q^2 + M_W^2)^2} \times
  \mbox{\hspace{23mm}} \\
  \left[ Y_+ {F_2}^\CC + Y_- {x F_3}^\CC - \ y^2 F_L^\CC  \right]
\end{eqnarray*}
\begin{eqnarray*}
  (F_2^\CC,xF_3^\CC) = x \left[ (u + c + \bar{d} + \bar{s}, u + c -
    \bar{d} - \bar{s}) \right],
\end{eqnarray*}
however, for electrons the $W$ couples only to positively charged
quarks. The reduced cross section is defined as $\tilde{\sigma}^\CC =
\frac{d^2 \sigma^\CC} {dx\,dQ^2} \times \frac{4 \pi x}{G_F^2}
\frac{(Q^2 + M_W^2)^2}{M_W^4}$.

\section{\boldmath Inclusive $\phi(1020)$-meson production}
$\phi$ mesons in DIS can be either formed directly from $s$ quarks
inside the proton or indirectly from $s$ quarks produced via BGF or in
the hadronization process. Generally, the cross section for indirect
production is much larger than that for the direct one, rendering the
measurement of the $s$-quark content of the proton rather difficult.
However, for direct $\phi$ production the variable $x_p(\phi) =
2p(\phi) / Q$ clusters around 1, where $p(\phi)$ is the $\phi$
momentum in the Breit frame. Therefore, at high $x_p(\phi)$ inclusive
$\phi$ production is sensitive to the $s$-quark content of the proton.

The $e^+p$ data set used was recorded during the run periods 1995--97
at a center of mass energy of $300 \gev$ and corresponds to an
integrated luminosity of ${\cal L} = 45.0 \pb^{-1}$. The cross section
was measured in the range $10 < Q^2 < 100 \gev^2$ and $2 \times
10^{-4} < x < 10^{-2}$.  The preliminary cross section is $\sigma(e^+p
\rightarrow e^+ \phi p) = 0.506 \pm 0.022 \nb$ \cite{phi_paper}.
\begin{figure}[t]
  \begin{center}
  \includegraphics[width=6.5cm,height=6cm]{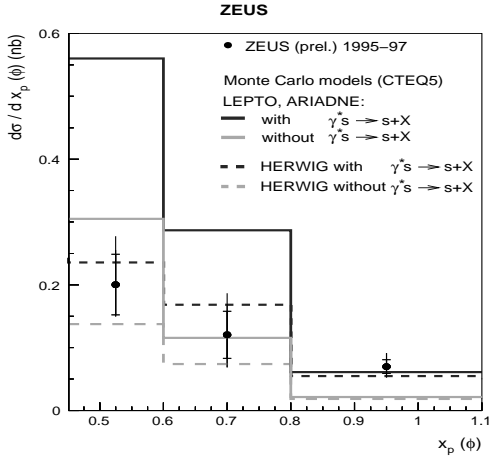}
  \vspace*{-9mm}
  \caption{$d \sigma / dx_p$ as a function of $x_p$. The dark
    (light) lines correspond to predictions using LEPTO/ARIADNE and
    HERWIG with (without) inclusion of an $s$-quark content of the
    proton.}
  \label{fig-phi}
  \end{center}
\end{figure}
In Fig.~\ref{fig-phi} for $x_p \lesssim 0.8$, where the indirect
contribution to $\phi$ production is expected to dominate and
theoretical predictions are somewhat uncertain, large differences
between the various models can be observed. However, for $x_p > 0.8$
the differences among models with $s$-quark content on the one hand
and among those without $s$-quark content on the other hand are small
compared to the measurement error. Here, models featuring an $s$-quark
content of the proton are clearly  required by the data.

\section{\boldmath $F_2$ from initial-state radiative events}
The acceptance of the ZEUS main detector for scattered electrons is
limited towards small scattering angles to $\theta^\star \gtrsim
4^\circ$ ($Q^2 \gtrsim 2 \gev^2$). One possibility to measure $F_2$ at
even lower $Q^2$ values is to use initial-state radiative (ISR) events
where a photon was radiated from the incoming lepton. Due to the
energy carried by the photon, $E_\gamma$, $\sqrt{s}$ of the reaction
is reduced by a factor $(E_e - E_\gamma) / E_e$. As a consequence,
electrons with low $Q^2$ are scattered into the main detector.

The $F_2$ measurement \cite{ISR_paper} was performed on a small
fraction of the 1996 $e^+p$ data set (${\cal L} = 3.8 \pb^{-1}$). It
covers the range $0.3 < Q^2 < 22 \gev^2$, $1 \times 10^{-5} < x < 3
\times 10^{-2}$ and fills the gap in the $F_2$ coverage of the
kinematic plane around $1 \gev^2$ and $x > 3 \times 10^{-4}$.
\begin{figure}[t]
  \begin{center}
  \includegraphics[width=6.5cm]{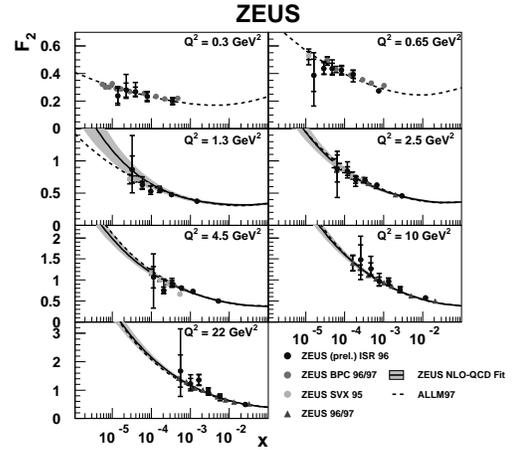}
  \vspace*{-9mm}
  \caption{$F_2$ as a function of $x$ in different $Q^2$ bins. In
    addition to the ISR data also former measurements are shown
    together with theoretical predictions.}
  \label{fig-isr}
  \end{center}
\end{figure}
Fig.~\ref{fig-isr} shows that the ISR $F_2$ values agree well with
other measurements both in the lower and upper $Q^2$ region.
Theoretical predictions based on the ZEUS NLO-QCD fit are in good
agreement with the ISR data for $Q^2> 2.5 \gev^2$. For lower $Q^2$ a
Regge inspired fit (ALLM97) yields a good description of the data. ISR
events will allow a direct measurements of $F_L$ in the future.

\section{High-$Q^2$ NC cross sections from $e^-p$ DIS}
Electromagnetic interactions are invariant under $P$ and $C$
transformations and hence, $\sigma^\NC(e^-p) \approx \sigma^\NC(e^+p)$
for $Q^2 \ll M_Z^2$ ($\gamma$-only exchange), where $M_Z$ is the mass
of the $Z$ boson. However, weak interactions do not preserve $P$ and
$C$ but only approximately $CP$. As a consequence $\sigma^\NC(e^-p) >
\sigma^\NC(e^+p)$ for $Q^2 \gtrsim M_Z^2$, where in addition to the
$\gamma$ exchange also the $Z$ exchange gives a considerable
contribution to the cross section. The parity violating terms of
$\sigma^\NC(e^\pm p)$ are combined in the structure function $xF_3$,
whereas those in $F_2$ are invariant under $P$. A comparison of
$e^-p$ and $e^+p$ cross sections yields a direct test of the
electroweak sector of the Standard Model. 

\begin{figure}[t]
  \begin{center}
  \includegraphics[width=6.5cm]{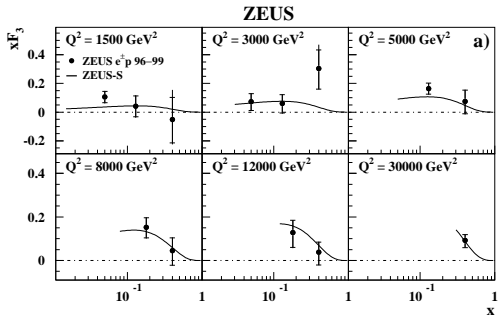}
  \vspace*{-9mm}
  \caption{$xF_3$ as a function of $x$ in $Q^2$ bins.}
  \label{fig-nc2}
  \end{center}
\end{figure}

The $e^-p$ data set used was recorded in 1998--99 (${\cal L} = 15.9
\pb^{-1}$). The NC cross sections were measured in the range $185 < Q^2 <
50\,000 \gev^2$ and $0.0037 < x < 0.75$ \cite{NC_paper}.
\begin{figure}[t]
  \begin{center}
    \includegraphics[width=6.5cm,height=8.5cm]{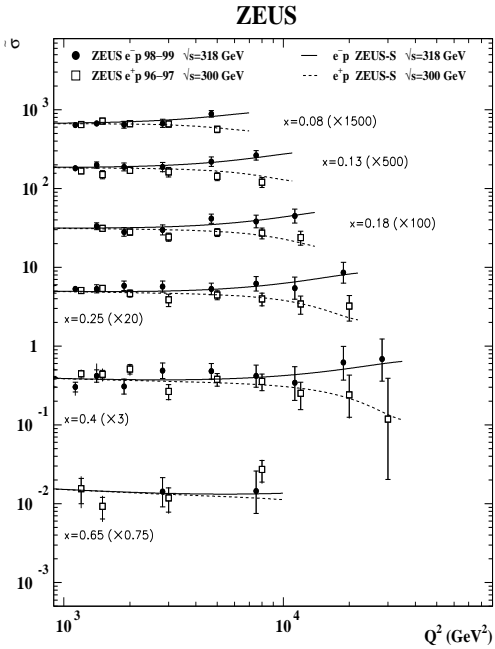}
    \vspace*{-9mm}
    \caption{The reduced cross section $\tilde{\sigma}$ as a function of
      $Q^2$ in different $x$ bins.}
    \label{fig-nc1}
  \end{center}
\end{figure}
Figure~\ref{fig-nc1} shows good agreement between the theory computed
with the ZEUS fit to the 1996--97 data (ZEUS-S) and the data, where
only points with $Q^2 > 1\,000 \gev^2$ are shown. Also plotted is the
$e^+p$ measurement from 1996--97 data based on $30 \pb^{-1}$. A
comparison of $e^-p$ and $e^+p$ cross sections shows an
increase of the former and a decrease of the latter as expected from the $Z$
exchange. The structure function $xF_3$ extracted from these data is shown in
Fig.~\ref{fig-nc2} and is compared to the prediction from the ZEUS-S NLO fit.
The measurement yields $xF_3 \neq 0$ and the data is well described by
the theory using ZEUS-S. The large errors of the measurement are
dominated by the low $e^-p$ statistics.

\section{High-$Q^2$ CC cross sections from $e^-p$ DIS}
In contrast to NC reactions CC reactions are purely weak and exhibit
therefore a more direct dependence on the weak parameters. Also, due
to the charge-selective coupling of the $W$ boson, measuring CC cross
sections yields information about the flavor content of the proton.

The $e^-p$ data set used for the measurement \cite{CC_paper} was
recorded in 1998--99 (${\cal L} = 16.4 \pb^{-1}$). The measurement was
performed for $280 < Q^2 < 30\,000 \gev^2$ and $0.015 < x < 0.42$.
\begin{figure}[t]
  \begin{center}
  \includegraphics[width=6.5cm,height=7.5cm]{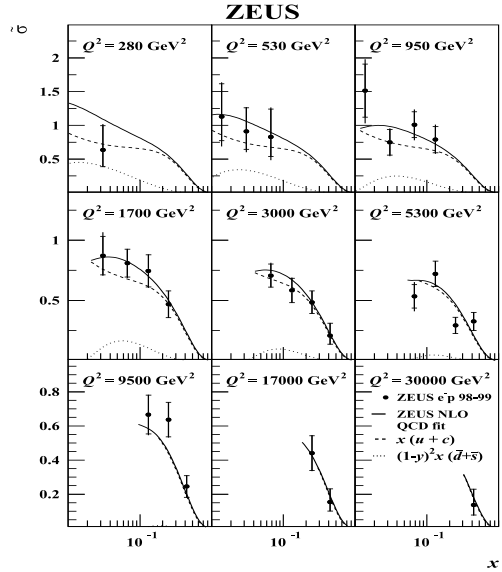}
  \vspace*{-9mm}
  \caption{Reduced charged current cross section as a function of $x$
    in different $Q^2$ bins. Shown are the measured and expected cross
    sections and the contributions from $u + c$ and $\bar{d} + \bar{s}$
    quarks.}
  \label{fig-cc}
  \end{center}
\end{figure}
Figure~\ref{fig-cc} shows good agreement between data and predictions
using ZEUS-S. At high $x$ the cross section is dominated by the $u$
(valence) quarks and a measurement in this region allows a direct
determination of the $u$ quark PDF of the proton. At low $x$ the
contribution from the $(\bar{d} + \bar{s})$ sea is also small due to
the helicity structure of the reaction. As the CC cross section
directly depends on the mass of the $W$ boson, $M_W$, a fit of $M_W$
to $d\sigma / dQ^2 \propto M_W^4 / \left( Q^2 + M_W^2 \right)^2$ was
performed. The fit yielded $M_W = 80.3 \pm 2.1 \, (\stat) \pm 1.2 \,
(\syst) \pm 1 \, (\PDF) \gev$. A comparison with the fit to the $e^+p$
data ($M_W =
81.4\strut^{+2.7}_{-2.6}\strut^{+2.0}_{-2.0}\strut^{+3.3}_{-3.0}
\gev$) shows good agreement within errors. The systematic and PDF
uncertainties are much smaller for the more recent measurement. In
case of the systematic uncertainty, this is due to a much improved
understanding of the energy scale uncertainty. On the other hand, the
$u$-quark PDF, dominating the $e^-p$ cross section, is much better
known than the $d$-quark PDF which mainly determines the $e^+p$ cross
section. The $M_W$ values, measured in the space-like region, are in
good agreement with and complementary to the time-like measurements
from LEP and Tevatron.

\section{NLO QCD analysis of data on DIS}
\begin{figure}[t]
  \begin{center}
  \includegraphics[width=6.5cm,height=7.6cm]{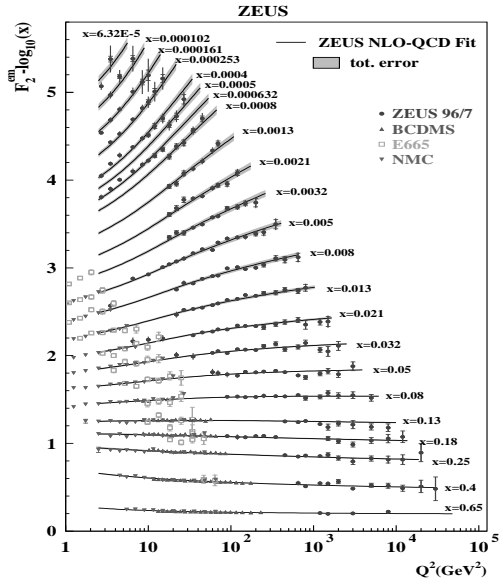}
  \vspace*{-9mm}
  \caption{$F_2$ as a function of $Q^2$ in bins of $x$.}
  \label{fig-fit1}
  \end{center}
\end{figure}

The global NLO QCD fit to ZEUS data \cite{NLO_fit}, including
measurements until 1997, and to fixed target data from BCDMS, NMC,
E665 and CCFR uses the full information on point-to-point correlations
of systematic uncertainties. The data used cover the kinematic range
$2.5 < Q^2 < 30\,000 \gev^2$ and $6.3 \times 10^{-5} < x < 0.65$.  The
fit, displayed in Fig.~\ref{fig-fit1}, gives an excellent description
of the data down to $Q^2 \approx 0.8 \gev^2$. The new precise ZEUS
data yield information on the gluon distribution and the quark
density at low $x$. A fit to the data with $\alpha_s$ free gives
$\alpha_s = 0.1166 \pm 0.0053$. The uncertainty of the PDFs is
dominated by the correlated systematics. The ZEUS fit is compatible
with both the MRST2001 and CTEQ6M parameterizations.

\begin{figure}[t]
  \begin{center}
    \includegraphics[height=3.2cm]{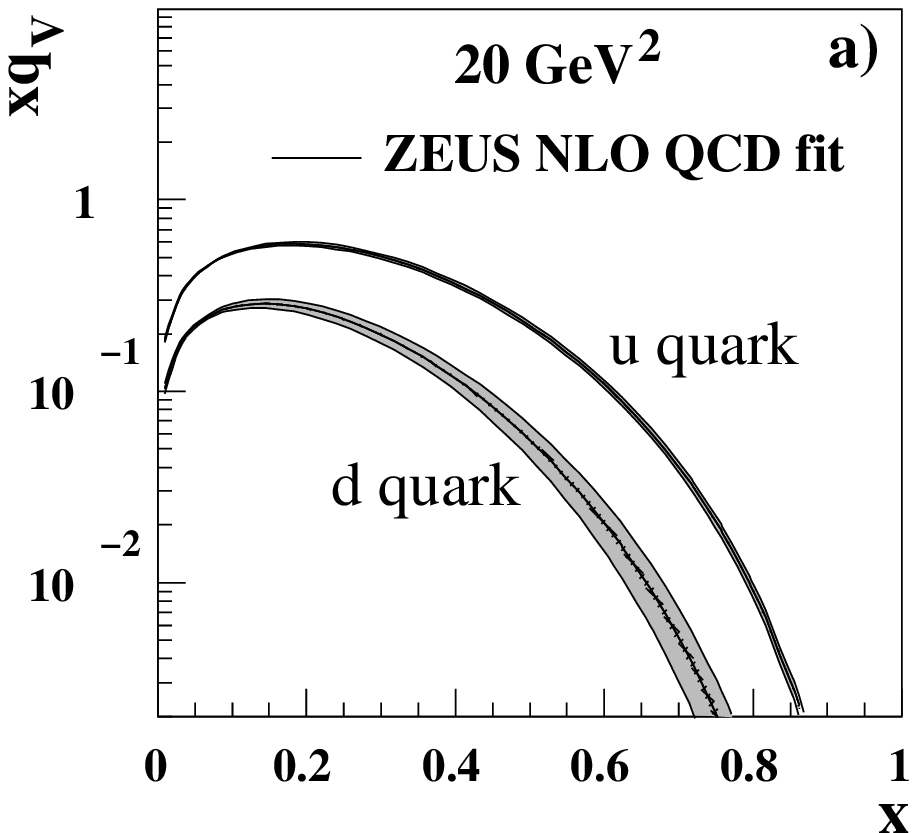}
    \includegraphics[height=3.2cm]{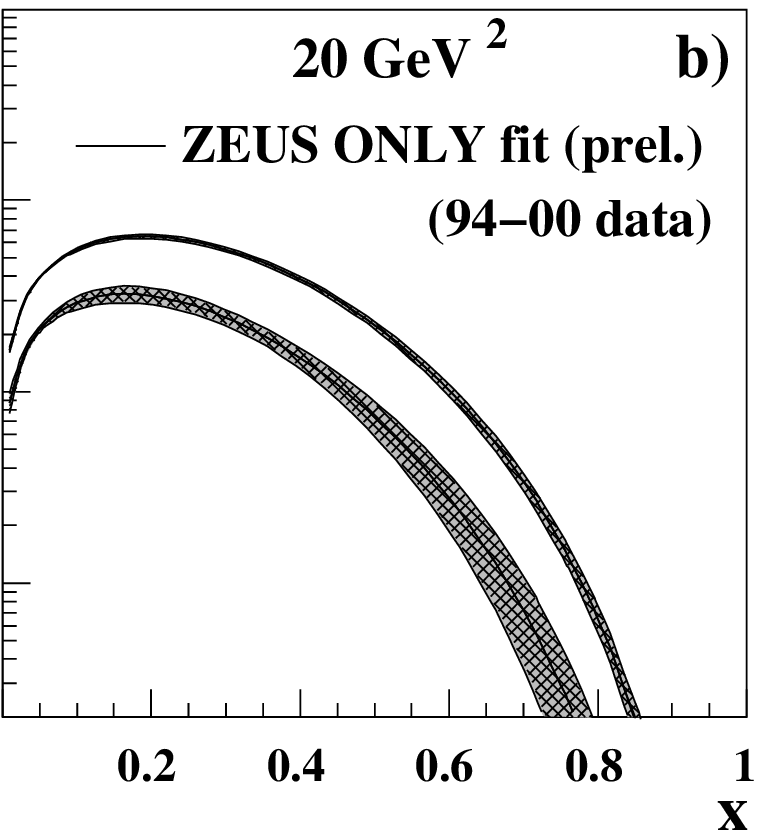}
  \vspace*{-9mm}
  \caption{Valence quark distributions from a) the standard ZEUS fit and
    b) the fit to ZEUS data only, including the 1998--2000 data.}
  \label{fig-fit2}
  \end{center}
\end{figure}

A fit to ZEUS only data including the newest data from 1998--2000,
displayed in Fig.~\ref{fig-fit2}\,b) yields similar uncertainties on
the valence quark PDFs as the standard fit, shown in
Fig.~\ref{fig-fit2}\,a), which includes the fixed target data.

\section{Summary}
Recent precision DIS data from ZEUS complete the coverage of a large
area of the kinematic plane spanning 6 orders of magnitude in $x$ and
$Q^2$. All ZEUS results are fitted consistently by NLO QCD and yield
precise PDFs.

\begin{thebibliography}{9}
\bibitem{phi_paper} ZEUS Coll., Abstract 850, $31^\text{st}$
  International Conference on High Energy Physics, Amsterdam,
  Netherlands, 2002.
\bibitem{ISR_paper} ZEUS Coll., Abstract 771, $31^\text{st}$
  International Conference on High Energy Physics, Amsterdam,
  Netherlands, 2002.
\bibitem{NC_paper} ZEUS Coll., S.~Chekanov et al., Preprint
  DESY-02-113, 2002, ICHEP abstract 766.
\bibitem{CC_paper} ZEUS Coll., S.~Chekanov et al., Phys.~Lett.~{\bf
    B~539}, 197 (2002), ICHEP abstract 763.
\bibitem{NLO_fit}  ZEUS Coll., S.~Chekanov et al., Preprint
  DESY-02-105, 2002, ICHEP abstract 765.
\end{thebibliography}
\end{document}